# Synchronizing Audio-Visual Film Stimuli in Unity (version 5.5.1f1): Game Engines as a Tool for Research.

Authors: Javier Sanz (1), Andreas Wulff-Abramsson (2), Luis Emilio Bruni (2), Carlos Aguilar (1) and Lydia Sánchez (1)

(1) Department of biblioteconomia, documentació i comunicació, Biblioteconomia i documentació, Universitat de Barcelona, Barcelona, Spain.
(2) Department of Architecture, Design and Media Technology, Aalborg University, Copenhagen, Denmark.

**Summary**
Unity is a software specifically designed for the development of video games. However, due to its programming possibilities and the polyvalence of its architecture, it can prove to be a versatile tool for stimuli presentation in research experiments. Nevertheless, it also has some limitations and conditions that need to be taken into account to ensure optimal performance in particular experimental situations. Such is the case if we want to use it in an experimental design that includes the acquisition of biometric signals synchronized with the broadcasting of video and audio in real time. In the present paper, we analyse how Unity (version 5.5.1f1) reacts in one such experimental design that requires the execution of audio-visual material. From the analysis of an experimental procedure in which the video was executed following the standard software specifications, we have detected the following problems desynchronization between the emission of the video and the audio; desynchronization between the temporary counter and the video; a delay in the execution of the screenshot; and – depending on the encoding of the video – a bad fluency in the video playback, which even though it maintains the total playback time, it causes Unity to freeze frames and proceed to compensate with little temporary jumps in the video. Finally, having detected all the problems, a compensation and verification process is designed to be able to work with audio-visual material in Unity (version 5.5.1f1) in an accurate way. We present a protocol for checks and compensations that allows solving these problems to ensure the execution of robust experiments in terms of reliability.

**KEYWORDS**
Unity, video, delay, movie texture, experimental design, methodology

## 1. INTRODUCTION

To capture and analyse the reactions of the spectator to certain audio-visual events, we need a system that allows us to synchronize different measurements together with different chosen audio-visual clips. Unity is a software mainly for videogame and simulation development [1-4]. However, this tool can also be very useful in research, because it allows to schedule actions and events, and to capture virtual behavioural data synchronized with the process of the experiment. It allows a great versatility of programming, including the possibility of working with video and synchronizing video signals with external sources of measurements, making it an appropriate environment for the development of experiments. The versatility of Unity allows parallel processing of actions in videogames, like for example record and store eye-tracking information while analysing user game behaviour. Similarly, Unity also allows network-based synchronization with external programs such as Matlab to relate electroencephalogram

(EEG) signals to in game events [5,6]. Similar procedures have been used to study the EEG of subjects exposed to content different from video games, such as for example exposition to abstract and figurative artistic paintings [7]. However, the use of video in Unity requires an elaborate and precise methodology in order to avoid synchronization errors.

Integration of video in Unity has been used to perform different research experiments [8-10]. To use videos in Unity a "movie texture" has to be employed, which applies the video as if it was a texture on a well illuminated two-dimensional element in Unity's three-dimensional space [11]. This procedure allows to integrate a video in Unity, however, this integration is not optimal, because it implies a high consumption of computer resources by Unity that can influence the correct functioning of the system. Given that the synchronization of psychophysiological or behavioural recordings with the viewing of the video is critical in research, in order to record the measurements in synchrony with the progress of the video we must verify that the flow of the system does not introduce delays or any desynchronization between the different processes executed by Unity.

In order to check how video execution affects the consumption of resources in the Unity system and if this consumption of resources can produce errors in the data record of an investigation, we have designed a set of tests to verify the reliability and accuracy offered by Unity when playing videos in the way described in previous literature, that is, in the mode of film texture and with video files converted to the format .ogg Theora [12].

## 2. METHODOLOGY

The objective of the experimental design was the synchronization of the EEG and the ocular scanner records of subjects exposed to films with the events pre-established on those films. During the development of the experiment it was detected that the Unity software could be producing erroneous data in the records of the experiment, so it was decided to design a test to verify the reliability of the system. Once synchrony problems were detected, solutions were developed to register the data correctly. The methodology, despite being designed for a concrete experiment and applied on Unity 5.5.1f1, could be applied in other experiments involving video files as well as in any other versions of Unity. In many investigations, the temporal precision of data records requires absolute robustness, this methodology is applied to check the coherence of the data record and to be able to compensate it in case of detecting delays in the records. The test methodology consists of two steps. The first step, analyses the playback quality of the video in Unity at different configurations to determine the maximum quality of the video format supported by the system. Once this has been determined, the second step analyses the reliability of the data recording process (saved in a text file), by comparing it with the progress of the audio-visual experience.

### 2.1. Step 1

The first part of the methodology is focused on checking whether the frame flow and the quality of the video managed through Unity is stable and fluent. The tests are done with two different scenes from different films. The final scene of the movie *Bonnie &*

*Clyde* [13] and the scene of Julian's death in *Children of men* [14]. The video is prepared following the ideal conditions as indicated in the Unity documentation; i.e.: encoded with Theora and below of 8356 kbps, which is the maximum supported natively, considering that working with video in High Quality consumes many more resources than in Normal Quality [12]. We start a source file with codec H264 at a frequency of 25fps progressive, with a resolution of 1280x720 and square pixel (1.0) with variable bitrate VBR Min 5 Mbps - Max 7 Mbps and audio with AAC codec in Stereo with a frequency of 48kHz and a bitrate of 192Kbps. We convert the source file to OGG with a maximum image bitrate of 7000kbps. We activate the option Soft Target, which makes the bitrate control less strict and usually offers better quality combined with Two pass encoding. For the audio we apply a bitrate of 192kbps and a frequency of 48kHz (it was done using the following program [15]). As we have a video with a maximum bitrate of 7000kbp, after this recoding we set the video quality of Unity to 0.9, which corresponds to a video processing capacity of 7535kbps.

A fluency test is performed on this video to know if the frame rate reproduced in Unity remains constant and stable during playback. To verify this, a frame counter is superimposed onto the video and the video counter reproduced by Unity is recorded with a video camera. A Panasonic model HC-V700 camera was used, which allowed us to shoot at 50fps. It is important to be able to shoot with a frame rate that is at least double as high as 25, since we can have a double exposition (figure 1) due to a lag between the camera's shutter frequency and the video frequency in Unity. It is due to the shutter speed of the camera that a camera frame could start its capture in the middle of a frame and end it in the middle of the next.

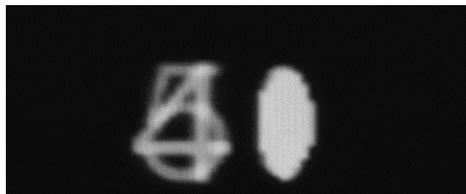

**Figure 1**     Screenshot of a double exposition in the frame counter

The verification of the correct running of the video in Unity can be deduced from a constant and stable advancement of a frame counter filmed by the camera. Cases in which jumps in the counter appear, or some numbers remain on the screen longer than what their frequency determines, can be considered a sign of poor fluency in the playback of the video by Unity.

**2.2. Step 2**

Once the adequate compression of the video has been checked, we proceed to perform reliability tests of the data recording and synchronization system. For this purpuse, we programmed a temporal counter in Unity that starts running just before the video is loaded. A controller is then activated to capture the frames that will be tagged by the time counter runningin Unity . To make the frames easily recognizable, we programmed the temporal marks in the transitions between shots (cuts) and at the beginning of the movie fragment. In addition to this, we kept the frame counter superimposed to the videofeed during the execution, which was captured by the video camera.

To check the synchronicity of the audio track with the video, the sound output is connected in line to the video camera adjusting the audio level to be equal to the original sound. Then, we compared the original clip with the video that captured the playback in Unity and its frame counter. The comparisson can be made with a normal editing program (in our case Premiere 2017).

This second step allows us to compare the temporal marker of the captured frames, the sound and the text file generated by Unity ,which ultimately enabled us to find any synchronization mistakes. This process rigouosly studies the perfect temporal execution of the video in Unity and the synchronicity between the different programmed actions. This is important to ensure if we are registering, for example,  eye tracking or electroencephalogram data simultaniously. If they are not perfectly synched, the programmed temporal markers are faulty, and if the audio does not run in synchrony with the video the percieved audio visual experience will be poor, thus affecting the outcome of the experiment.

## 2.3. The Unity Program

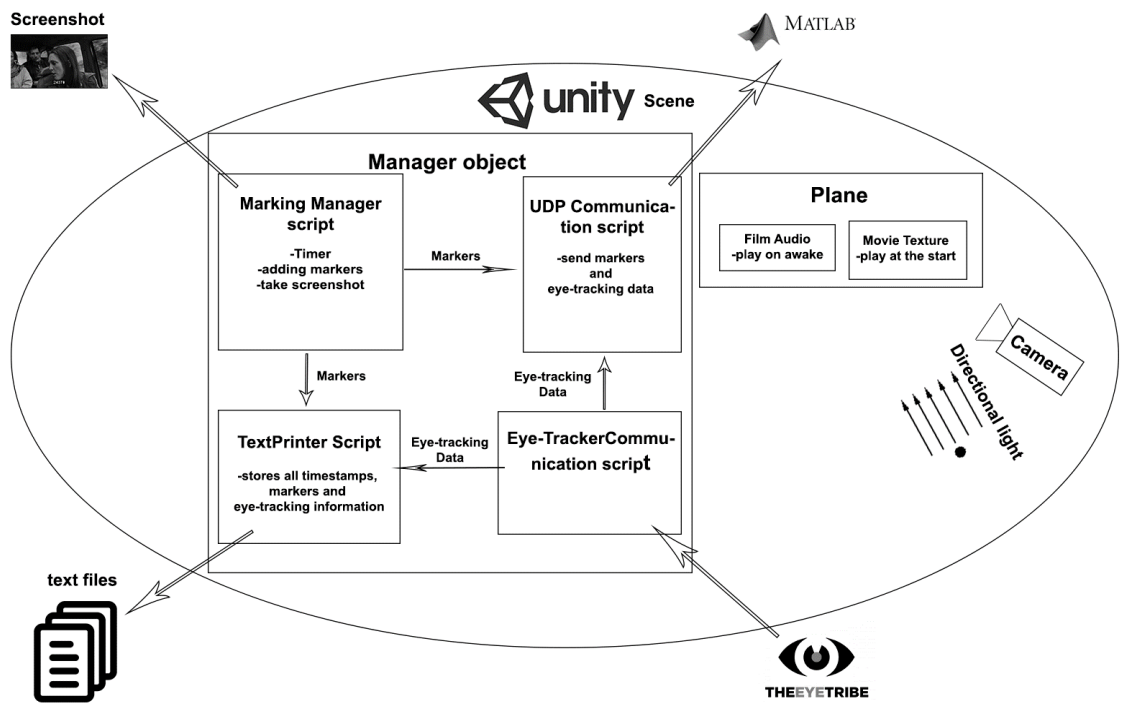

**Figure 2**  The building blocks used in the Unity scene. To the right the Plane on which the movie texture is attached is shown. The Plane is illuminated by a directional light and it is captured by an in-unity camera entity. To the left is the manager object. This object serves as a time tracker, marker manager and data communication entity to other outside programs

To make a scene in Unity (version 5.5.1f1), which is able to playback a video, the following elements are needed: a plane, a camera, a directional light and a manager object (see Figure 2,  right side). The plane is the surface on which both a material contaning the video texture and the audio file from the video are attached. Both parts of the stimuli are instructed to play when the scene is loaded. The light in the scene is a directional light and is pointing  at the plane in order to illuminate it as uniformly as possible. The Camera is directed at the plane as well, situated so the plane covers the whole viewing plane.

The manager object is an invisible object containing and running all the scripts necessary to run the experiement: A marking manager script, a UDP communication script, a textPrinter script and a eye-trackerCommunication script (see Figure 2, left side).

The marking manager script keeps track of time, which for the sake of synchronization is started at -0.251793, as it takes unity that amount of seconds to start the film. When this timer reaches a certain specified time for when there is a cut in the film, the script does several things: it instructs Unity to take a screenshot with the title (number of captions, minutes, seconds, frames), and it sends marker information to the textprinter script and the UDP communication script. When the timer reaches the length of the film clip the scene is changed to scene contaning another film clip.

The purpose of the UDP communication script is to establish a networking communication to programs outside Unity e.g. MATLAB. This serves the purpose of synchronizing the markers created in Unity with other datastreams e.g. physiological data recorded through MATLAB. Additionally, the UDP connection can also be used to transfer and synchronize other data such as Eye-tracking data.

The textPrinter script constantly receieves time and event marking data from the manager script, and together with eye-tracker data from the eye-trackerCommunication script, it fills up a list of all the received data. The data are all associated to specific time stamps, which are ordered in the list. At the end of the film clip the list of data is saved as a text file for post experiment inspection.

The last script of importance for the setup is the eye-trackerCommunication script. The purpose of this is to receive and communicate the eye-tracker data from e.g. an Eye-Tribe devise, which needs a TCP networking paradigm to get access to its data.

## 3. RESULTS

With this methodology, we can proceed to check how Unity plays back audio-visual content.

### 3.1. Step 1

The tests performed as step 1 show irregular fluency in the video playback. By observing the videorecorded frame-counter we can notice how some frames last longer than they should as well as the occurrence of jumps in the frame counter sequence. It can be seen that the duration of the clip remains the same as in the original.The fact that these frames are played for a longer time does not mean that the video slows down- Unity compensates the jams with small temporal jumps, thus skipping some frames. This test leads us to conclude that the playback of videos embedded as a texture in Unity respects the length of the fragment, but presents problems to gurantee an adequate fluency of the content.

Since the playback quality is of paramount importance, different converters to Theora codec were tried: VLC, Theora Converter.NET, Simple Theora Encoder and various other online alternatives. The converter that produced the best results was Theora Converter.net in terms of preserving visual quality and the smoothness of the playback in Unity. Therefore, it was decided to utilizeTheora Converter.net as the converter for multiple conversion tests.

### 3.1.1. Theora codification

Theora is a video encoding based on the VP3 codec. Theora allows a block-based motion compensation, a free-form variable bitrate (VBR), a minimum block size of 8x8, a flexible encoding of entropy, a subsampling format of 4:2:0, 4:2:2, and 4:4:4, 8 bits per pixel per channel colour, multiple frames of reference (frames) and intra frames (I-frames in MPEG), inter frames (P-frames in MPEG), but not B-frames (in MPEG4 ASP, AVC) [16]. Theora Converter.Net gives also the options to activate Soft Target and Two pass encoding. In addition to being able to specify the maximum bitrate, these options were not available in all the free converters.

Different coding configurations were tested using Theora Converter.Net in automatic and manual configurations. It was tested with two H264 1080p source files. One with variable bitrate VBR with a maximum bitrate of 8000kbps and another with constant bitrate CBR of 8000kbps. It was concluded that the best results in Unity arose from the variable bitrate CBR and with the configuration in Theora Converter.Net set to Soft Target and Two Pass encoding activated. However, there were still traces of fluency problems. Nevertheless, modifying the bitrate of the source file before converting the file could get rid of the problem, as the bitrate-specific data depends on the source file.

### 3.1.2. Source video file

Once the configuration of the converter to Theora was established, we tried alterations to the source file utilizing Premiere 2017 [17]. The previous work with Theora enabled us to achieve better results when coding the source file, thus knowing how to make Theora Converter.Net perform the least amount of transformations as possible. The framerate was changed to 24 and 30 but showed no improvements in Unity, as Unity plays the file back at 25 fps. Therefore the importance of the framerate was discarded. Different source bitrates were tested as well between 5000 and 8000 kbps, but the resulting videos did not significantly improve their fluency in the Unity execution.

Once the variation of the bitrate was ruled out as the main factor affecting the quality of the fluency in the Unity playback, tests were performed on video resolution. Reducing the size of the frame to 1080 x 720 pixels provided better results, diminishing the jumps and the delays of the playback in Unity. Despite this, the behaviour of the video was still not optimal, so we decided to resort to a configuration similar that of a standard DVD, with an SD resolution of 576 x 480 pixels in progressive format. The 480p video configuration played fluently in Unity, but it is considered the minimum threshold to maintain an acceptable quality of content reproduction. Therefore, the following tests revolved exclusively around reducing the bitrate, applying signal limiters, and signal compressors.

## 3.2. Step 2

In order to guarantee a smooth reproduction of the video in Unity, the source file has to be in H.264 PAL DV (16: 9), with resolution 576x480 at 25fps progressive, with a variable bitrate VBR 2 passes at 2-4 Mbps, and Main profile and appearance D1DV PAL 1,094. In addition, the options of signal limiters and compressors offered by Premiere 2017 should be applied. More specifically, the file has a video limiter with Intelligent limit (-30 - 130) on axis reduction and the reduction method is Compress all (shadow 64-10, Highlight 192-10). The sound was set to stereo AAC at a frequency of 48000Hz and a bitrate of 192Kbps. Afterwards, the conversion process to Theora was configured with a maximum bitrate of 4000kbps, with soft target and two pass encoding activated. The sound was set at a bitrate of 192kbps and a frequency of 48000Hz. When loading the video in Unity, the quality was set to 0.475, which is equivalent to 4000 Kbps. Once the most optimal coding of the reproduction of the video was determined we proceeded with the synchronization and data recording tests. The first results showed desynchronization between the audio and the video when played in Unity.

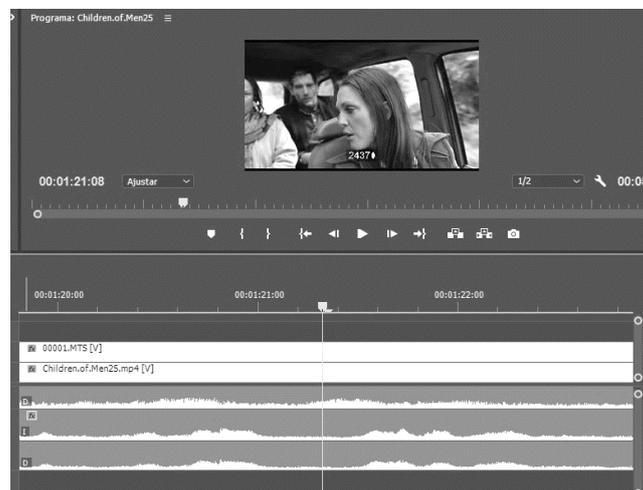

**Figure 3**    Screenshot of the frame code in the source video and the audio delay

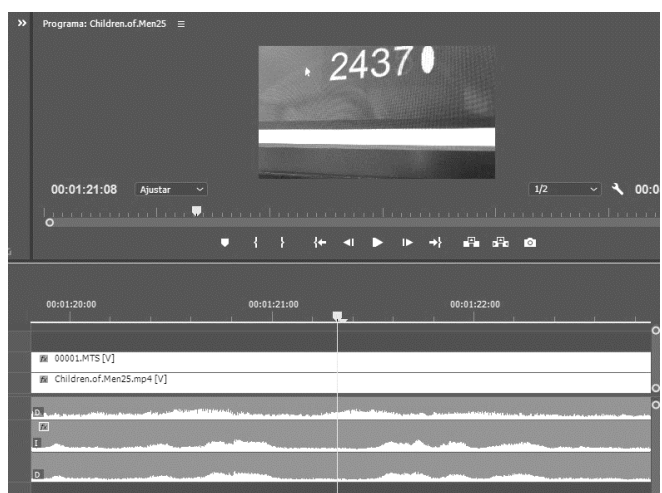

**Figure 3**    Screenshot of the frame code filmed and the audio delay

When we compared the playback with the embedded frame code (recorded by our videocamera) with the original clip in the edition software, the sound tracks did not match. We could observe desynchronization between the audio and the video when reproduced in Unity.When playing videos in Unity, you have to load video and audio separately in sequential orders. This procedure introduces a delay in the reproduction of the audio with respect to the video. The same synchrony test between audio and video was applied in different clips from different movies and the delay resulted to be variable between 5 and 6 frames, but in all cases it remains constant throughout the video. Specifically, it was applied in addition to the scenes of *Bonnie & Clyde* [13] and *Children of men* [14] in scenes of *On the Waterfront* [18], *The searchers* [19] and *Whiplash* [20].

In addition to verifying the synchrony, we want to check that when Unity records information associated with an event placed in a specific time of the video, the recorded data corresponds precisely to the corresponding frame of the video event. To do this, we recorded the frame number that Unity is programmed to capture in a text file and compared it to the captured frames. The captured frames overwrote the frame number with the corresponding frame counter. By comparing both recordings, a constant delay is prevailing between the capture of the frame and and the text file containing the programming order. If we convert the time recorded in the frame counter and compare it with the value of the frame counter that was overwritten in the captured pictures, the values did not match. Of the 79 captures programmed for Unity registration, 69.6% are delayed by 3 frames, 17.7% by 4 frames and 12.7% by 2 frames. This delay between 2 and 4 frames could have the same explanation as the audio delay. The time counter starts executing after the playback instruction of the video, but Unity introduces a delay in the playback of the video, which could be the reason why the counter and the video were desynchronized. Due to this lag, introduced by the video management, we must add an error of +/- 1 frame, thus a maximum error margin of +/- 0.04 seconds as the video is played at 25fps,achieving thereby a configuration where we get maximum synchrony.

Observing the time register file generated by Unity (Table 1), it can be observed that a constant frequency of appearing frames emerged when a screen capture function was called. This implies that the screen capture needs a certain time to finish, which introduces a delay in the captured image.

**Table 1** Time recorded in the register file

| Registered Time in seconds | 5.840649 | 5.85721 | 5.873774 | 5.890343 | 5.906908 | **6.039409** | 6.055976 | 6.074334 | 6.090903 |
|---|---|---|---|---|---|---|---|---|---|
| Time increased between one frame and the next in seconds | 0,016561 | 0,016564 | 0,016569 | 0,016565 | **0,132501** | 0,016567 | 0,018358 | 0,016569 | 0,016565 |
| Increased time converted in frames | 0,414025 | 0,4141 | 0,414225 | 0,414125 | **3,312525** | 0,414175 | 0,45895 | 0,414225 | 0,414125 |

Taking into account that the file is played at 25 fps, the frame capture process is done with a delay of 2.89 frames. It is therefore estimated that the frame capture generated a 2 frames delayed with respect to what it should capture.

### 3.3. Error correction

The correct synchronization between the playback of the audio-visual stimuli in Unity and the recording of data from external sensors is influenced by the delay the process of video and audio reproduction generates when executed in Unity. There are two processes that introduce delays. The first one concerns the correct reproduction of the video in Unity and the second the data recording. Solving the issue requires the correct synchronization of the audio-visual input with the recording of the signals coming from the peripherals managed by Unity. This requires two independent procedures.

a) Compensate the desynchronization between visual and audio. In order to correct the desynchronization between audio and visual when played in Unity, it is necessary to compensate the detected delay by modifying the video that will be played back. For this, the video that was going to be played was created by advancing the sound track the number of frames that we previously calculated. This video alteration allows the reproduction of audio and video in Unity to be synchronized. The only way to solve this problem is creating a source video with a desynchronization that will be compensated by Unity when the audio-visual is played back.

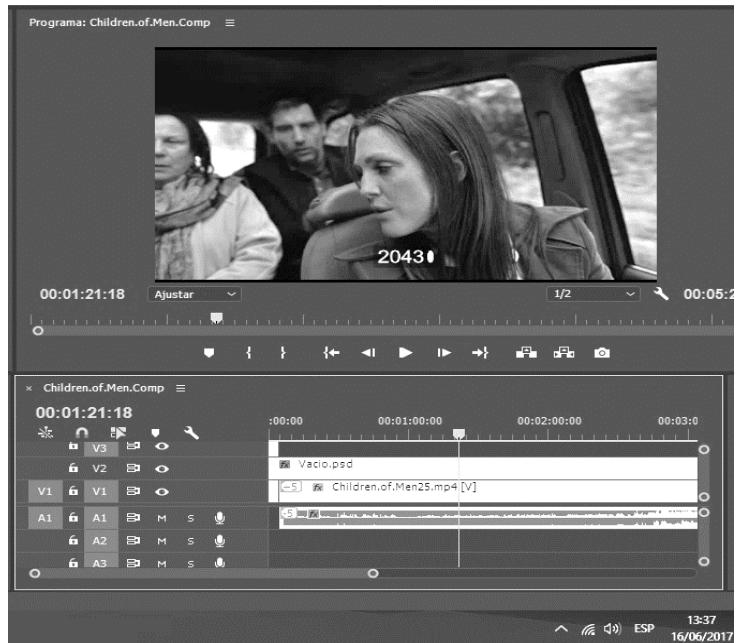

**Figure 4** Screenshot of the audio delay fixed in the source video

In this way, the video with the advanced audio will not suffer the delay introduced by Unity resulting in a synchronized playback. We re-run the synchrony test again and observed that Unity played back the video in perfect synchrony.

b) Compensate the delay entered in the event record. Unlike the problem of synchronization between video and audio, programming a compensation for the delay in the execution of Unity did not pass the synchrony tests, so the error, despite remaining constant throughout the execution, starts with a delay variability that is impossible to predict. For this reason, the compensation must be made on the recorded data, once the lag introduced by the registration process has been verified in each specific case. The delay introduced by the capturing process is calculated first by comparing the captured frame with the overwritten frame counter, and then by calculating the difference between the time saved in the text file for that capture and the time in the video, which corresponds to the captured frame. We used this delay value to compensate all the temporal values in the text files that Unity created during its execution. Through this process we cannot compensate the Unity counter in its execution due the +/-1 frame of delay range, but we managed to synchronize the data recordings with the audio-visual input.

## 4. CONCLUSSION

To conclude we would like to summarize the proposed methodology but first it is necessary to clarify that it may not be universally applicable. The method must be adapted for each experiment by repeating the procedures for the concrete experiment. Furthermore, to avoid other computer related synchronization biases, it is also necessary to deactivate all the unnecessary software and block the maintenance activities of the computer on which Unity is running. The proposed methodology is divided into three blocks. It is important to follow these blocks in the order indicated because the results of the subsequent blocks are conditioned by the configuration of the processes in the previous blocks.
1- Define the encoding parameters for the video.
2- Compensate the audio.
3- Compensate the temporal text file.

The first and second block must be implemented before the execution of the experiment, while the third block is performed afterwards on the recordings from the experiment.

### 4.1. Define the encoding parameters for the video

The first step is to define the conversion software that will be used. To do this the codec, format and bitrate characteristics, supported by our version of Unity, were collected in order to perform standard conversions for all of our chosen clips. Once we have converted our battery of clips, we load them into Unity and perform the fluency test with a video camera. This will help optimize the configurations of the conversion to Theora software. Once the converter was reconfigured, clips with different conversion characteristics to Theora were created. We returned to the fluency test to elucidate if further optimization of the video files were needed.

The next step is to define the characteristics of the video clip before they are converted to Theora. For this we tested an ideal Theora configuration for Unity by performing the fluency test with the video camera. If we did not achieve an acceptable fluency we modified the features of the videos to approach a lighter reproduction file. If the first alteration of the video files did not produce adequate fluency, we repeated the previous step. After several tests, the clips will be of the best quality that Unity can reproduce in an acceptable way. In this step we must have in mind that certain parameter alterations can provoke an alteration to the conversion configuration in Theora, thus requiring returning to the first step before continuing.

### 4.2. Compensate the audio

The compensation of the audio desynchronization is done in six steps.
a- Video with frame counter overwritten.
b- Filming Unity as it plays the video with the frame counter, and while the audio is passed to the camera.
c- Calculate the delay between the audio and the video.
d- Create a new video file with compensated audio.
e- filming the compensated video with time counter overwritten played by Unity.
f- Check that Unity reproduces video and audio synchronously.

The first step is to add a frame counter to the video file. In this way we will be able to work with precision during the comparison between the clip and the filming of the playback in Unity. The second step is to film the playback while passing the sound from the computer to the camera. Once step (b) is done, we can compare the filmed track with the original clip in an editing program and adjust them by following the frame counter. This allows us to calculate the delay that occurred in the play back in Unity between the video and the sound. When the delay is known, we advance the audio in the original clip in such a way that the delay estimated between the execution of the video and the sound in Unity will be compensated.

Once we have the video file with the compensated audio we re-test it as we have done in the step (b). At this time, we should observe that when synchronizing the video files in the editing software by following the frame counter, the audio must coincide between the filmed and the original clip. In case they do not coincide, it should maximum be with a difference of +/- 1 frame. This can be caused by the filming frequency of the video camera and the playback of the video not matching up. In this case we return to step (e). If the delay was maintained, we deduced that the problem was not in the filming of step (f), but in step (b). Thereafter we readjust the desynchronization calculated in step (c) adding or subtracting a frame change and go back to step (d). We repeated the steps as many times as necessary until the audio and video played in Unity are synchronized.

### 4.3. Compensate the temporal record text file

The compensation of the temporal record text file has 5 steps. One of them is done before performing the experiment and the remaining are performed on the recordings after each experiment.
a- Create synchronization marks and programming its screenshot.
b- Define the time difference between the screenshot and the programmed temporal registry that ordered the capture.
c- Calculate the delay of the capture.
d- define the delay between the time counter and the video playback in Unity.
e- Compensate the Unity registry text file.

To understand and compensate for the temporal delay between the screen capture and the text file with the temporal information about the screenshots, a black screen of minimum 10 frames in the beginning of the clips with a superimposed frame counter must be added to the clips. Additionally, a synchronization mark should be programmed to take a screenshot at the beginning of the clip that will show the black frame and the counter. An alternative option is to add the synchronicity mark at the end of the video. One can also make two synchronization marks, one at the beginning and one at the end in order to enhance the reliability. Once the synchronization mark has been added to the system, the steps previous to the experiment are concluded. The remaining steps concern the recorded data text file.

Once the experiment is carried out, we convert the time of the text file to frames and compare it with the synchronization mark's frame counter in order to calculate the first delay value. Then we compared the time recorded in the text file time for the synchronization mark screenshot with the next temporal recording. This will define the delay caused by the screenshot. Once these two values are obtained we know the real delay that exists between the Unity timer and the video playback. Finally, we apply this correction to all the temporal data recorded in the text file from the experiment. Thereby, the text file will be synchronized with the original video input.

**REFERENCES**

1.  *Cities: Skylines.* (2015). [Video Game]. Developed by Colossal Order.
2.  *Escape Plan.* (2012). [Video Game]. Developed by Fun Bits Interactive.
3.  *Firewatch.* (2016). [Video Game]. Developed by Campo Santo.
4.  *Rust.* (2013). [Video Game]. Developed by Facepunch Studios.
5.  Bruni, L. E., Bacevicuite, S., & Arief, M.. *Narrative Cognition in Interactive Systems: Suspense-Surprise and the P300 ERP Component*. Interactive Storytelling (2014, Novmeber 3), Lecture Notes in Computer Science book series Vol. 8832, 164-175.
6.  Wulff-Jensen, A., & Bruni, L. *Evaluating ANN efficiency in recognizing EEG and Eye-Tracking Evoked Potentials in Visual-Game-Events.* 8th International Conference on Applied Human Factors and Ergonomics, Los Angeles, EE.UU., 2017. Springer International Publishing.
7.  Baceviciute, S., Bruni, L., Burelli, P., & Wulff-Jensen, A. *Differences in Cognitive Processing When Appreciating Figurative and Abstract Art Can Be Detected by Integrating EEG and Eye-Tracking Data.* 24th Conference of the International Association of Empirical Aesthetics, Vienna, Austria, 2016.
8.  Genovese, A., Craig Jr., C., & Calle, S. *3ME-A 3D Music Experience* (2010). ResearchGate.
9.  Larkee, C., & LaDisa, J. *An efficient approach to playback of stereoscopic videos using a wide field-of-view.* Electronic Imaging (2015), 1-6.
10. Ramachandrappa, A. *Panoramic 360◦ videos in virtual reality using two lenses and a mobile phone.* Doctoral dissertation, University of Illinois, Illinois, EE.UU., 2015.
    Link: https://ideals.illinois.edu/handle/2142/89067
11. Thorn, A. *Unity Animation Essentials.* Packt Publishing, 2015, Birmingham, England.
12. Unity Technologies. (2017). *Unity User Manual (2017.3).* Retrieved 2017, from https://docs.unity3d.com/Manual/index.html
13. *Bonnie and Clyde.* (1967). [Motion Picture]. Directed by A. Penn.
14. *Children of Men.* (2006). [Motion Picture].Directed by A. Cuarón, A.
15. Ratkiley. *Theora Converter .NET.* Retrieved April 12, 2018, from https://sourceforge.net/projects/theoraconverter/
16. theora.org. *Theora Format Specification.* Retrieved 06 13, 2017, from: http://theora.org/doc/Theora.pdf

17. Adobe. *Premiere Pro CC 2017*. Retrieved April 13, 2018, from https://www.adobe.com/products/premiere.html
18. *On the Waterfront.* (1954). [Motion Picture]. Directed by E. Kazan.
19. *The Searchers.* (1956). [Motion Picture]. Directed by J. Ford.
20. *Whiplash.* (2014). [Motion Picture].Directed by D. Chazelle.